# Long-Period Oscillations of Sunspots Observed by SOHO/MDI

V.I. Efremov[1]   L.D. Parfinenko[1]
A.A. Solov'ev[1]   E.A. Kirichek[1]

Abstract
   We processed magnetograms that were obtained with the Michaelson Doppler Imager onboard the Solar and Heliospheric Observatory (SOHO/MDI). The results confirm the basic properties of long-period oscillations of sunspots that have previously been established and also reveal new properties. We show that the limiting (lowest) eigenmode of low-frequency oscillations of a sunspot as a whole is the mode with a period of 10 - 12 up to 32 - 35 hours (depending on the sunspot's magnetic- field strength). This mode is observed consistently throughout the observation period 5 - 7 days, but its amplitude is subject to quasi-cyclic changes, which are separated by about 1.5 - 2 days. As a result, the lower mode with periods of about 35 - 48 hours appears in the power spectrum of sunspot oscillations. But this lowest mode is apparently not an eigenmode of a sunspot because its period does not depend on the magnetic field of the sunspot. Perhaps, the mode reflects the quasi-periodic sunspot perturbations caused by supergranulation cells that surround it. We also analyze SOHO/MDI artifacts, which may affect the low-frequency power spectra of sunspots.

1. Introduction

The best-studied part of the oscillatory spectrum of sunspots is in the range of periods from three to five minutes (Leibacher and Stein, 1971; Beckers and Schultz, 1972). However, recent studies have revealed the low-frequency part of the spectrum, in which a sunspot oscillates slowly with periods from 30 - 40 minutes to dozens of hours, and where it clearly manifests itself as a localized solitary magnetic structure (Gelfreikh et al., 2006; Bakunina et al., 2009a, 2009b; Chorley et al., 2010; Efremov et al., 2007, 2009, 2010, 2012; Kallunki and Riehokainen, 2012; Smirnova et al., 2011, 2013a, 2013b).

---
[1] Central (Pulkovo) Astronomical Observatory,
St. Petersburg, Russia
Email: maklimur@gmail.com

Studying long-period oscillations of the solar atmosphere opens up new possibilities for probing deep layers of the Sun. But for a comprehensive study, which covers a sufficiently wide range of frequencies, data from space observatories are needed to record long homogeneous time series for different physical parameters.

In this article we analyze data obtained with the Michaelson Doppler Imager (MDI: Scherrer et al., 1995) instrument onboard the Solar and Heliospheric Observatory (SOHO: Domingo, Fleck, and Poland, 1995) regarding of the long-period variations of the sunspot magnetic- field strength.

In 2006 - 2007, we investigated the low-frequency oscillations of sunspots by means of line-of-sight velocity variations. Measurements of Doppler shifts of spectral lines of a sunspot provide direct evidence of quasi-periodic movements of the gas in a sunspot along the line-of-sight. To detect this effect, we used long series of digital spectrograms of sunspots, from four to eight hours, obtained at the ground-based solar telescope of the Pulkovo Observatory. The solar spectrum in a range 649.379 - 649.965 nm was recorded with the Pulkovo telescope's digital camera size of 22.2 × 14.8 mm and contains seven solar spectral lines, formed at heights from 190 to 535 km above $\tau_{500} = 1$, as well as one telluric line used to control the influence of the terrestrial atmosphere. Recording Doppler shifts of the selected lines allowed us to study the long-period oscillations of sunspots and surrounding magnetic elements with periods of 40 - 45, 60 - 80, and 135 - 170 minutes (Efremov et al., 2007). It was found that a sunspot as a whole moving with these periods slowly oscillates in the vertical direction, which causes quasi-periodic vertical displacements. Apparently, the balance of pressure between the sunspot and its surroundings is maintained due to the relative slowness of the process. As a consequence, the vertical displacements of a sunspot change the radius of its magnetic flux tube, and (due to magnetic flux conservation) the average magnitude of the magnetic field also changes, over a cross section. It varies in time with the same period as the line-of-sight velocities. In a previous article (Efremov et al., 2009) we investigated the change in amplitude of the 80-minute, low-frequency mode in a sunspot using Doppler shifts of spectral lines with different formation heights. It was shown that the amplitude of the plasma oscillation mode decreases rapidly with height: distinctly detectable at 200 km, it becomes bearly perceptible at a heights of about 500 km; see Figure 1 in Efremov et al. (2010). This fundamentally distinguishes the low-frequency oscillations from short period (three -five minute) fluctuations. Their amplitude in the observable layers of a sunspot does not decrease with height, but even tends to increase. It is also necessary to point out that the oscillations of low-frequency modes persist in sunspots during the entire observation time. It was reliably established that low-frequency oscillations are completely absent from the telluric line and a quiet photosphere that is free from magnetic elements (Efremov et al., 2007). This means that oscillations of this type are seen exclusively in solar magnetic structures.

Based on the theoretical model of a "shallow sunspot" (Solov'ev and Kirichek, 2008, 2009), one can derive the possible existence of modes with an even lower frequency in the oscillatory spectrum of sunspots. However, because the ground-based observations are naturally limited by their duration, one cannot detect variations of line-of-sight velocities (or magnetic- field strength) in sunspots with periods exceeding three - four hours. Therefore, we used magnetograms from archived MDI data (Scherrer at al., 1995).

It has been established that the low-frequency modes detected previously that were expected according to the theory are clearly seen on space magnetograms (Efremov et al., 2009, 2010, 2012). These independent data obtained above the terrestrial atmosphere establish the physical reality and solar origin of the long-period oscillations of sunspots.

While processing space-based magnetograms with a duration of dozens of hours, we detected two new low-frequency modes with periods of about 250 and 480 minutes (at a confidence level σ > 3 ); their amplitudes increase monotonically with the period (Efremov et al., 2010).

In this article, we continue to study low-frequency oscillations of sunspots over very long (up to 150 hours) time intervals. We use magnetograms of sunspots obtained with a cadence of 1 and 96 minutes from MDI.

2. Processing and Measurement Techniques

Data from SOHO/MDI and the Helioseismic and Magnetic Imager (HMI: Scherrer et al., 2012) onboard the Solar Dynamics Observatory (SDO: Pesnell, Thompson, and Chamberlin, 2012) provide time sequences of full-disk Sun magnetograms (soi.stanford.edu/data/). To track the temporal variation of the sunspot magnetic field, one needs to select magnetic- field strength from each magnetogram at strictly the same place in the sunspot throughout the series of multi-day observations. Therefore, a thorough correction of the data is required for the rotation of the Sun and the proper motion of the sunspots.

We selected a small part of the magnetogram, so that the sunspot remains within it during the entire observation time. Handling such a relatively small area instead of processing the full solar disk reduces the required computer resources and time significantly.

To form a time series of magnetic- field strength in a sunspot, we used the extreme-value method (Efremov et al., 2010). The time series of magnetic field are formed by an extreme value of the pixel in the sunspot. These extreme pixels are located according to the topology of a particular sunspot, which is not necessarily located at its geometrical center.

3. Results of Processing the MDI Magnetograms

To study the low-frequency modes of sunspot oscillations, we processed the following magnetograms with a 96-minute cadence (the initial day of a series is specified; in some sequences it was possible to process more than one sunspot): 12 February 1998, 27 April 1998, 15 June 1998, 1 November 1998, 24 March 1999, 16 May 1999, 11 June 1999, 15 July 1999, 19 September 1999, 26 May 2000, 22 June 2000, 24 March 2001, 29 March 2002, 12 August 2002, 20 August 2002, 19 September 2002, 5 October 2002, 3 March 2003, 11 March 2003, 21 January 2005, 14 April 2005, 31 August 2005, 14 November 2005, 24 November 2005, 7 June 2006, 29 June 2006, 1 October 2006, and 4 June 2007.

The duration of some series is up to 360 hours. Taking into account possible edge effects, we used the shorter series of magnetograms (from four to six days long), when the sunspot was not too far away from the disk center.

In addition, we used a long (up to 117 hours) series of magnetograms with a one-minute cadence. This was done to increase the spectral resolution and compare the spectral characteristics of magnetic- field oscillations with results of ground-based observations. Long continuous series of magnetograms such as this (without gaps and defects) are relatively rare. Gaps in data considerably complicate the analysis in Fourier space (Stahn et al., 2008). To achieve reliable results, we used only flawless series of magnetograms of the following dates: 24 September 1999, 9 October 1999, 14 July 2000, 23 July 2000, 22 November 2000, 9 September 2001, 31 March 2002, 27 July 2002, 27 October 2003, and 13 January 2005.

We applied a wavelet transformation to study variations of the sunspot magnetic field. Processing about 40 series of magnetograms yielded reliable and consistent results.

The time series of the extreme-values of a field [H] can be represented as $H(t) = H_{01} + H_{02} + h(t)$, where $H_{01}$ is the temporal change of the magnetic field of a sunspot due to projection effects [$\sim \cos(\theta)$], $H_{02}$ is the temporal change of an average magnetic eld due to some slowly temporal evolution of the sunspot (these two components have no periodic components and are combined in the concept of the "slow temporal trend"), and, finally, $h(t)$ is the fluctuating part of the series. This fluctuating part is the focus of our attention here. Therefore, to obtain a solution to our problem, we do not need the exact absolute value of the magnetic- field strength, which means, in turn, that our approach does not need to account for possible calibration errors.

3.1. Long Time-series with One-Minute Cadence

Figure 1a presents the processing results of a series with a total duration of 4096 minutes (68 hours) with a one-minute cadence.

The top panel of Figure 1a shows the part of the solar surface in the form of a narrow strip that contains the sunspot studied (NOAA AR 08706, 24 - 26 September 1999). The strength of the magnetic field in the sunspot is $H_Z (\Theta = 0) = 2750$ G. Below, the variation of magnetic field [$H_Z(t)$] during the observation is presented. Furthermore, the normalized time series of the magnetic field in the sunspot and its wavelet transformation are given. To the right, a global wavelet spectrum (i.e. the integrated characteristic of the oscillatory process for all observation periods) is presented.

This time series $h(t)$ has a zero average ($<h(t)> = 0$) and a dispersion (normalized variance of the process) equal to unity ($D[h(t)] = 1$), which is presented in dimensionless form.

For this time series, the level of confidence of the spectral components can be simply estimated by using Fisher's test (Koen, 1990).

In the global wavelet spectrum, two modes with periods of about 1000 and 520 minutes are clearly visible, with amplitudes that considerably exceed the 95 % confidence level. A similar study was performed for other long time series with one-minute cadence. Analyzing the collection of results obtained, we conclude that the limiting mode (the mode with the lowest frequency) of the magnetic-field oscillations in sunspots is that with a period of 720 - 2000 minutes (12 -32 hours).

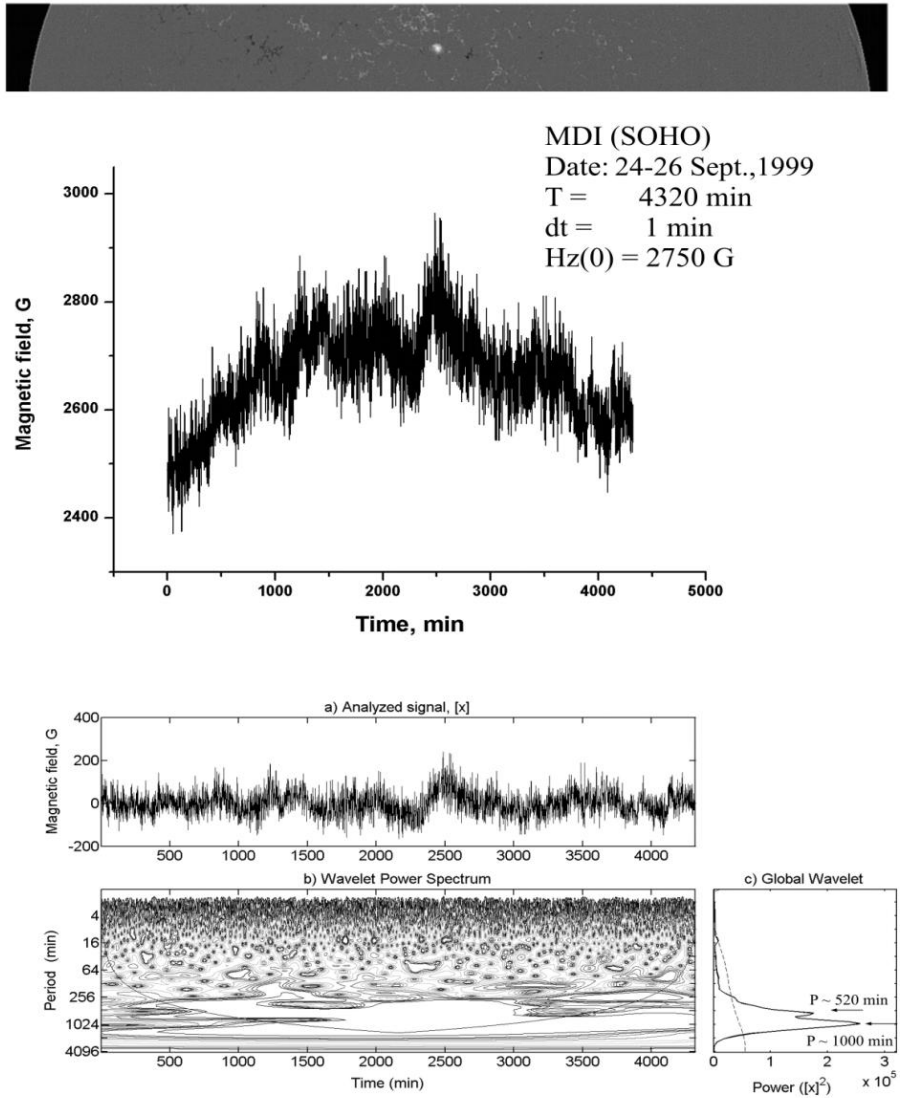

Figure 1. Top: narrow strip of the full-disk magnetogram, with the sunspot in NOAA AR 8706 (24 {26 September, 1999). Bottom: i) the temporal variation of the maximum magnetic- field strength [h(t)] in the sunspot; ii) time series, and iii) its wavelet transformation. Bottom-right: the global wavelet spectrum.

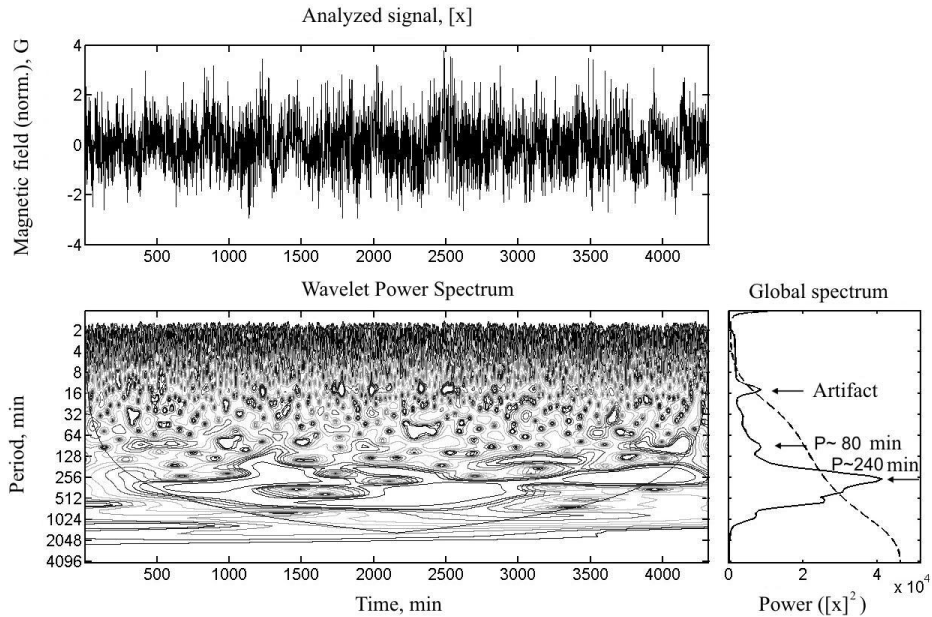

Figure 2. Time series for the sunspot in NOAA AR 08706 (same as in Figure 1a) (24 - 26 September 1999); two lowest-frequency components are filtered out.

We remark that this conclusion regarding the limiting mode was not obvious. Although it is not apparent in the example given in Figure 1a, in a power spectrum of sunspots there normally exists a strong mode with a period that is two to four times longer than the limiting value; it amounts to about 2100 - 2880 minutes (35 - 48 hours). At the beginning of the study, we assumed that just this second mode was the limiting one. The problem of this choice between the two lowest modes turned out to be significant; we discuss it in more detail in Section 4.

We resume discussing the above-mentioned time series (4096 minutes duration, with one-minute cadence) and present the results of the processing that was performed to reveal the high-frequency components of magnetic- field oscillations in sunspots. Figure 1b shows the power spectrum of the time series in which the two lowest modes are filtered out. The dominant mode is that with a period of about 240 minutes. There is another mode with a period of about 80 minutes.

Figure 1c presents the high-frequency part of a power spectrum. Here, the dominant mode has a period of 13 minutes. This reflects the presence of an artifact (false harmonic), that is caused by a feature of the readout from the pixel matrix of the MDI receiver. The rotation of the Sun moves the sunspot gradually over the pixel matrix. It takes exactly 13 minutes (the period of false mode) for it to move from one pixel to the next. This artifact does not reveal itself in a uniform magnetic field, but in a real sunspot umbra with a gradient of the magnetic- field strength, a small variation of the measured magnetic- field strength appears as the point of extreme field moves from one pixel to the next (Efremov et al., 2010).

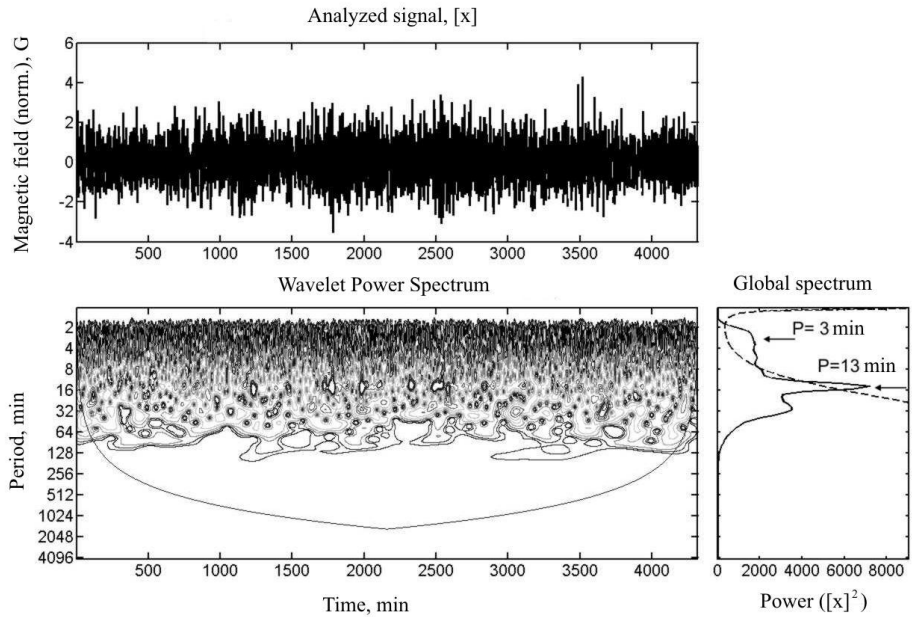

Figure 3. Time series for the sunspot in NOAA AR 08706 (the same as in Figure 1a) (24 - 26 September 1999) and its wavelet spectrum; the highest frequency components are shown only.

Thus, using the MDI data, we have found the well-known high-frequency modes as well as the low-frequency modes that were found previously using ground-based observations, with periods of 40, 80, 160, and 240 minutes (see Efremov et al., 2009, 2010). In additional, two new modes with periods of about 520 and 1000 minutes have been revealed.

Another time series that was investigated has a longer duration (94 hours) and the same high temporal resolution (one-minute cadence). The processing results are presented in Figure 2. The sunspot NOAA AR 9887 was observed from 31 March to 4 April 2002. The wavelet spectrum clearly shows the above-mentioned strong mode with a period of about 2100 minutes (35 hours).

It is important to emphasize that modes with periods longer than 35 - 50 hours were not found in any processed series (more than 40), while the length of the series (about 100 - 150 hours) and the technique applied allow one, in principal, to reveal modes with periods of up to 50 - 75 hours. This fact suggests the existence of a limiting eigenmode of sunspot oscillations. With this, we establish the property of the long-period sunspot oscillations. If the limiting

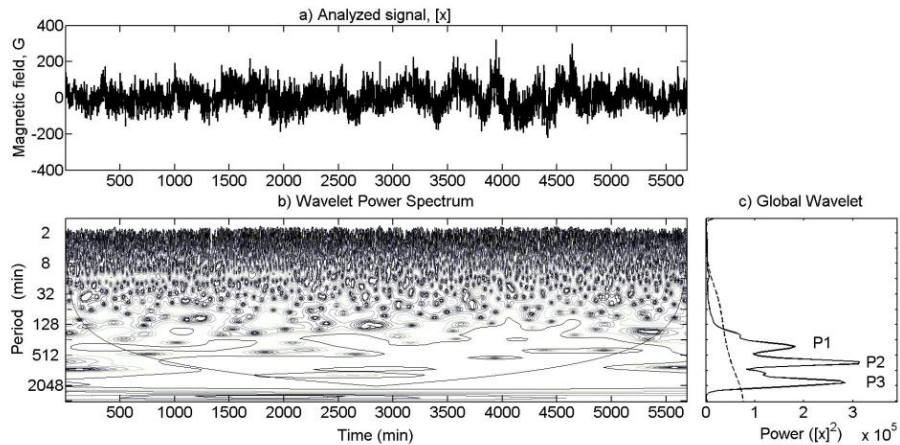

Figure 4. A typical wavelet spectrum for magnetograms with one-minute cadence of the sunspot in NOAA AR 09887 (31 March - 4 April 2002), Hz(0) = 3250 G. The low-frequency modes of magnetic- field variations and the modes with periods of about 400 (P1), 800 (P2), and 2000 minutes (P3) are visible.

mode did not exist, more lowest-frequency modes would appear when we treat a long time series.

### 3.2. Long Time Series with 96-Minute Cadence

We also considered a long series of magnetograms, recorded with a temporal cadence of 96 minutes. Data with this cadence do not allow one to study the oscillations with periods shorter than 180 minutes (three hours), but their advantage lies in their quantity: there are quite a few damaged magnetograms with a 96-minute cadence in the archives of MDI.

The results of a wavelet analysis of all processed time series look very similar. Normally, there is a mode with a period of 720 - 2000 minutes, and also a sufficiently strong lower-frequency mode with a period of 2100 - 2880 minutes. A typical wavelet for one of the 96-minute-cadence series (sunspot NOAA AR 10139, 5 October 2002) is shown in Figure 3.

### 4. Limiting Eigenmode

The main goal of the present work is the detection of the lowest eigenmode of sunspot oscillations. Two modes may qualify for this, as we mentioned above. The first one (M1) has a period of 720 - 1300 minutes (15 - 20 hours on average) and another one (M2) with periods between 2100 and 2880 minutes (40 hours on average).

In the beginning, it seemed quite natural to identify M2 as the limiting mode by regarding the higher harmonics as an overtone of this main mode. However,

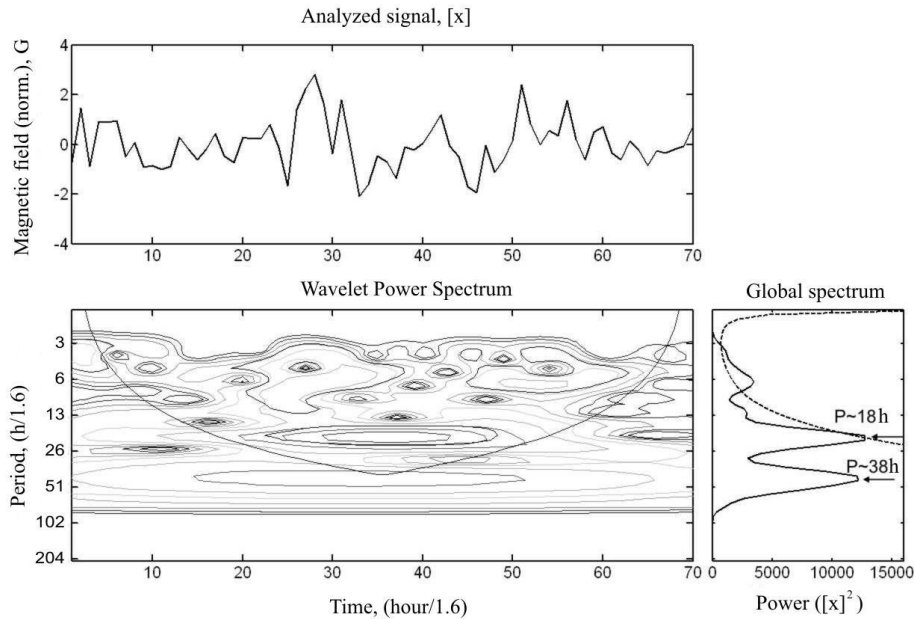

Figure 5. Typical wavelet spectrum for magnetograms with 96-minute cadence, of the sunspot in NOAA AR 10139, 5 October 2002, Hz(0) = 2700 G.

a careful analysis revealed fundamental differences in behavior between M1 and M2.

The first argument against considering M2 to be a limiting eigenmode is that there are some cases when the amplitude of this mode appeared to be lower then the amplitude of M1. Studying other spectral components, we have always observed that the amplitude of oscillations decreased with increasing frequency. The mode M2 does not obey this rule: its amplitude is often less then the M1, and sometimes is absent. This fact suggests that the M2 does not reflect the eigenmodes of sunspot being the harmonic of another nature.

The second aspect of the behavior of the M1 and M2 modes is that the wavelet analysis clearly reveals an unusual excitation character of M1 mode. The oscillations at these periods follow one another in the form of successive pulses (wave trains) and give rise to the quasi-stable oscillations pattern. Generally, the duration of the pulse is about two - three periods of M1. This temporal interval (30 - 45 hours) is, on one hand, close to the period of M2, and on the other hand it is the typical lifetime of supergranulation cells. (We recall that the characteristic lifetime of supergranulation cells is estimated to be about 15 - 30 hours. For some cells it can reach two and even four days; normally an interval of 35 hours is taken as the average lifetime for these structures (Leighton, Noyes and Simon, 1962).

The proximity of the oscillation period of M2 and the supergranulation life-time leads us to suppose that in the power spectra of sunspot oscillations we ob-serve the results of external quasi-periodic perturbations of a sunspot originating in the supergranulation cells that arise and break up close to a sunspot.

There is another argument in favor of this idea. The average horizontal size of supergranulation cells is 30 Mm (Leighton et al., 1962). Therefore, in addition to the coincidence of time scales, there is a coincidence of spatial scales of sunspots and supergranulation cells. Because of this coincidence of the temporal and spatial scales, resonant interactions might be expected.

Thus, the assumption that M2 is the result of quasi-periodic perturbations of a sunspot caused by its surroundings appears to be quite plausible. It suggests a different physical nature of M1 and M2.

Since M2 reflects the action of an external perturbing force on a sunspot, its period should not depend at all on the magnetic- field strength in a sunspot. In contrast, the period of M1, which is the fundamental eigenmode of the sunspot, should have a clear and specific dependence on the sunspot's magnetic- field strength.

Both experimental research and the theoretical model of the "shallow sunspot" (Solov'ev and Kirichek, 2008, 2009) show the special type of the dependence of the sunspot period eigenmodes on the magnetic- field strength of the sunspot. While the magnetic- field strength lies within an interval of 1000 and 2600 - 2700 G, the period of eigenmodes decreases with increasing field strength. In the interval between 2700 G and the maximal field strength, which is approximately 3400 - 3500 G, the period increases rapidly. At a magnetic- field strength of about 2600 - 2700 G, the period of the eigenmodes of the sunspot is at its minimum. The periods dependence of M1 and M2 on the sunspot magnetic field is presented in Figure 4.

Clearly, the periods of M1 and M2 depend in a different way on the sunspot magnetic field. M2 does not show any significant dependence on the magnetic field, while M1 reveals the expected type of period dependence on magnetic- field strength, as was predicted by the theoretical model.

Thus, summarizing our arguments, we conclude that the limiting eigenmode of sunspot oscillations is the mode M1 with a period of about 10 - 12 up to 32 - 35 hours, depending on the magnetic- field strength. If the magnetic- field strength approaches its maximum (about 3400 - 3500 G), the period of M1 can increase significantly and become close to the period of M2 (see Figure 4). Here one can expect resonance to take place when the sunspot oscillations are efficiently excited by an external force in a narrow band of frequencies (see Figure 5).

Another case, where the frequency of the eigenmode M1 differs significantly from the frequency of an external supergranulation mode M2 corresponds to a value of the magnetic- field strength of about 2600 - 2700 G. Then the situation is very different from the strong resonant one, and the power spectrum shows both the M1 and M2 harmonics (see example, Figure 3).

The fact that we are dealing specifically with the limiting mode M1 is shown well in Figures 3 and 5. The wavelet scaling is prolonged on purpose to 204 hours (see the vertical axes in the left-bottom panel in Figures 3 and 5). Although the length of the time series (144 hours) is sufficient to reliably detect the presence of the lowest modes with periods longer than 35 - 48 hours, these modes are completely absent from all processed series.

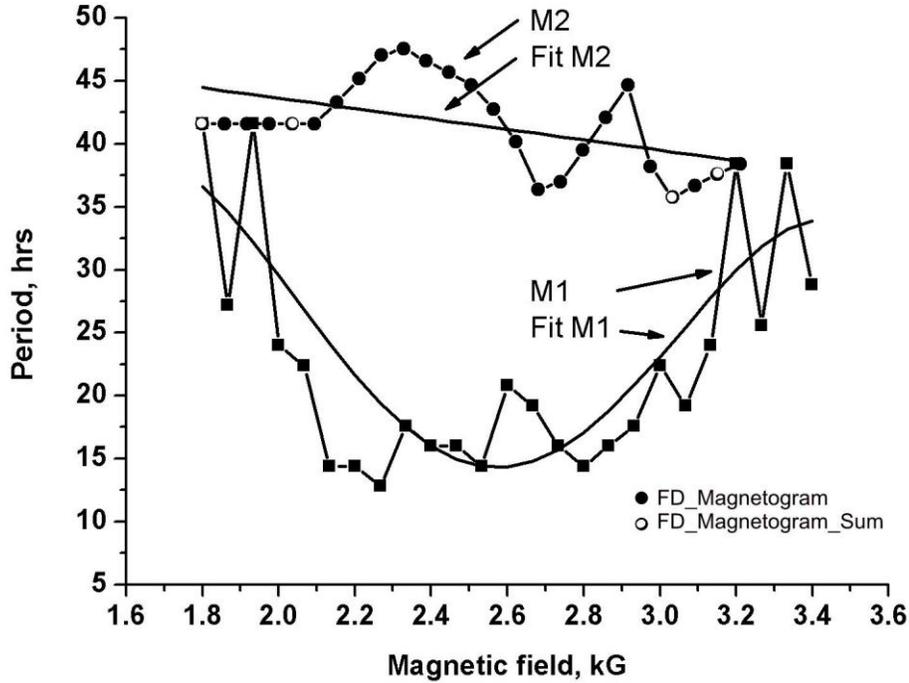

Figure 6. The dependence of the period of sunspot oscillations on magnetic-field strength $H_z$ ($\theta = 0$) for M1 (bottom curve) and M2 (top curve) modes. The eigenmode M1 approaches the external exciting mode of M2 at weak (about 1800 G) and strong (about 3300 G) magnetic fields.

## 5. SOHO/MDI Artifacts

Space data are free from the influence of the Earth's atmosphere. However, the spectral analysis of long time series of magnetograms may reveal a number of artifacts (false harmonics) that must be considered (see, for example, the study of artifacts of TRACE data: Yuan et al., 2011). Previously we have considered the artifacts that appear by processing MDI data in the low-frequency power spectra of sunspots due to discrete pixel structure of the MDI instruments matrix. This artifact was investigated by us in detail and referred to as the "P2P" effect (Efremov et al., 2010). Here we consider other possible artifacts that may be present in the MDI data.

### 5.1. MDI Magnetograms with One-minute Cadence

The data obtained with MDI contain artifacts that arise due to various instrumental effects (Raju et al., 2001; Liu et al., 2004). For example, the noise of the MDI shutter may cause a very faint artifact with a period of 5.33 hours (52.125

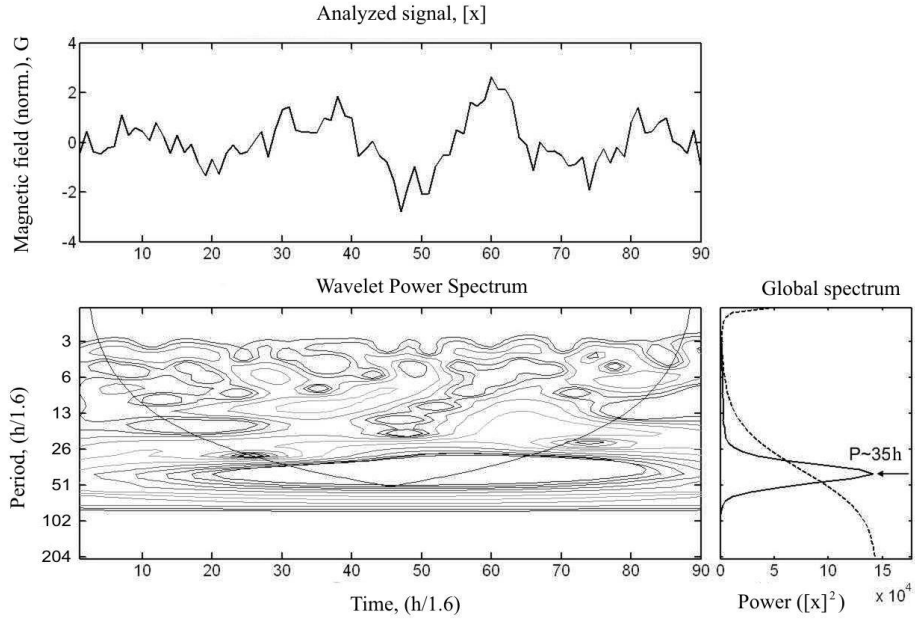

Figure 7. An example of resonant power spectrum. M2 excited by an external force coincides with M1. Sunspot in NOAA AR 10105 (10 September 2002), $H_z(0) = -3200$ G)

microhertz). In our results this period has never been found. But, theoretically, a very weak harmonic with periods of 320, 160, 80, 40, and 20 minute may appear in the signals of MDI. Another possible artifact of MDI data, caused by small changes in the position of the center of the solar image, also cannot affect our results, since we did not use the center coordinates to obtain the time series. As is known, there are some problems with the calibration of the MDI magnetograms, which can also lead to errors (Demidov and Balthasar, 2009). The overall effect of calibration reduces the field in the center of solar disk by 8 % and on the eastern limb by 30 % (soi.stanford.edu/magnetic/Lev1.8). As we mention in Section 3, the time series of the maximum values of a sunspot field [H] can be presented in the form $H(t) = H_{01} + H_{02} + h(t)$, where $h(t)$ is the fluctuating part of the series. We study this fluctuating part only. Therefore, we do not need the exact absolute value of the magnetic- field strength, and our approach is free from the calibration problems.

Long-duration time series (up to 140 hours) of MDI magnetograms with one-minute cadence often contain gaps and bad (defective) magnetograms. To eliminate the influence of these defects, we used only magnetograms of good quality that had no gaps.

These magnetograms can be successfully used for studying the low-frequency oscillations of sunspots.

### 5.2. MDI Magnetograms with 96-minute Cadence

There is a very large database of continuous series of MDI magnetograms with a cadence of 96 minutes. Due to the specific method of the telemetry transmission of these magnetograms to the Goldstone, California, USA station, artifacts with periods of 12 and 24 hours can arise in these data.

We have processed 30 series of 96-minute magnetograms with a duration of up to 360 hours. Our results suggest that no significant false harmonics with periods of 12 and 24 hours can be detected in the sunspots' power spectra. Probably, they are suppressed by more powerful harmonics of the sunspot eigenmodes. Sometimes the harmonics of eigenmodes can have a period of almost 12 hours, when the magnetic-field strength in the sunspot is about 2100 - 2200 G (see Figure 4).

We mainly used a series of FITS magnetograms, which are specified as OBS MODE = FD Magnetogram. Magnetograms indicated as OBS MODE = FD Magnetogram Sum appear to be more liable to artifacts caused by peculiarities of telemetry. However, we found no significant differences in these data types in the sense of the period of oscillation of the magnetic field. In Figure 4, the four points marked with white circles on the curve of M2 were obtained with OBS MODE = FD Magnetogram Sum. It is important to emphasize that these points do not deviate from the general curve constructed by means of OBS MODE = FD Magnetogram.

It is instructive to compare the power spectra of long-period oscillations of sunspots obtained from the magnetograms with a different cadence. Figure 6a shows the power spectra of the same spots obtained during 21 - 24 July 2004 from the time series of magnetograms with cadence of one minute (left) and 96 minutes (right). In both cases, the same low-frequency mode of sunspot oscillations with a period of about 33 hours is clearly registered. No signals of 12- and 24-hour periods are revealed.

We tested several synchronous series of magnetograms with cadences of 1 minute and of 96 minutes. An example for the time series of 8 - 10 October 1999 is presented in Figure 6b. To facilitate the comparison, the time series with one-minute cadence was reduced to a series with a 96-minute cadence. The independence of the low-frequency part of the sunspot power spectrum on the cadence of magnetograms is still clearly visible.

Thus, although the 12- and 24-hour artifacts, possibly, occur in a 96-minute series of MDI data, they are much less frequent than the real modes of sunspot oscillations, and therefore have no effect on our results.

It should also be noted that the magnetograms of 1- and 96-minute cadence for sunspots with magnetic-field strength greater than 3000 G are subject to saturation effects. Any magnetograms were rejected and not used in our study.

### 5.3. Independence of Oscillations in Widely Separated Sunspots

There is a simple opportunity to reveal the nature of the investigated oscillations: are they the real solar signals or artifacts? It is sufficient to investigate the degree of correlation of the magnetic-field oscillations in spatially separated sunspots that were observed simultaneously on the solar disk but do not belong to the same active region. The artifact (if it is there) should be manifested equally in all sunspots, regardless of their position on the disk. In this case, the oscillations in all sunspots visible on the solar disk should be strongly correlated. Conversely, if we observe genuine physical oscillations in widely separated spots that are not physically connected, these oscillations (even if their periods are similar because the magnetic fields in the spots are about the same) should be completely independent of each other.

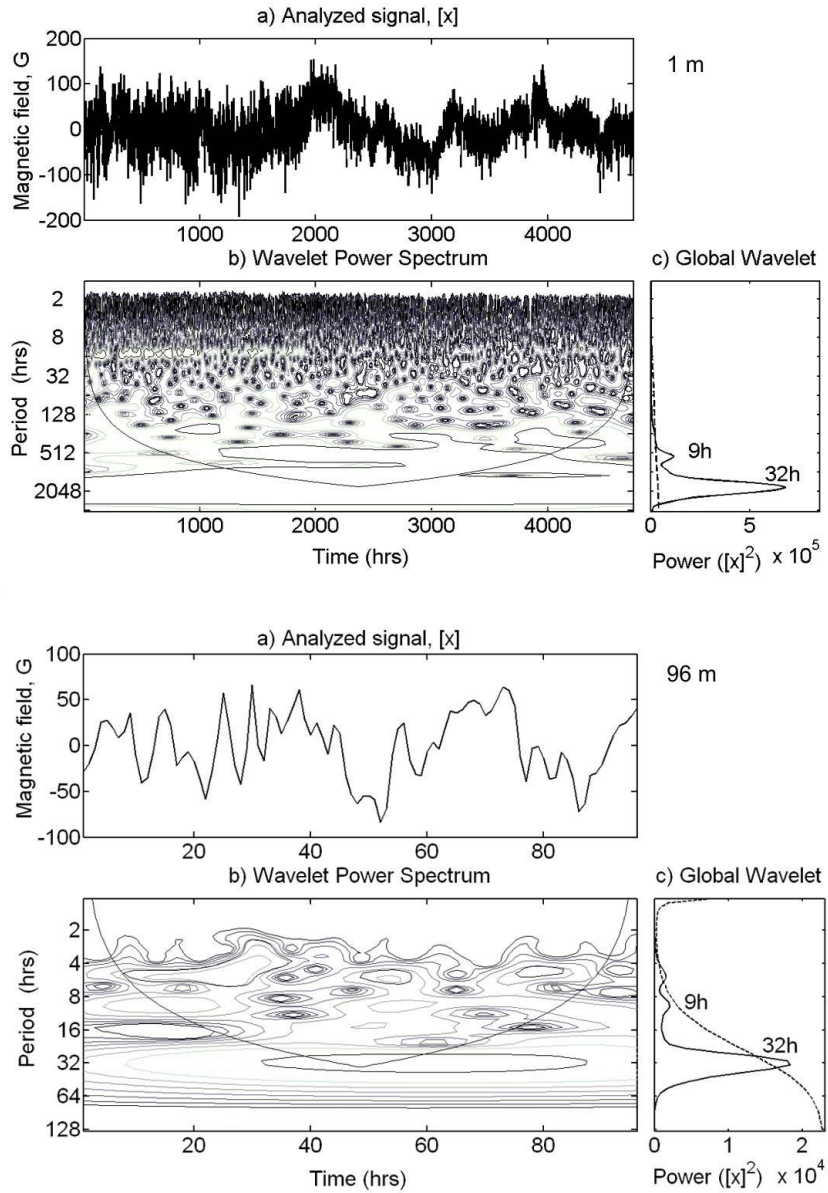

Figure 8. The independence of the low-frequency part of the sunspot power spectrum on the cadence of magnetograms (21 - 24 July 2004).

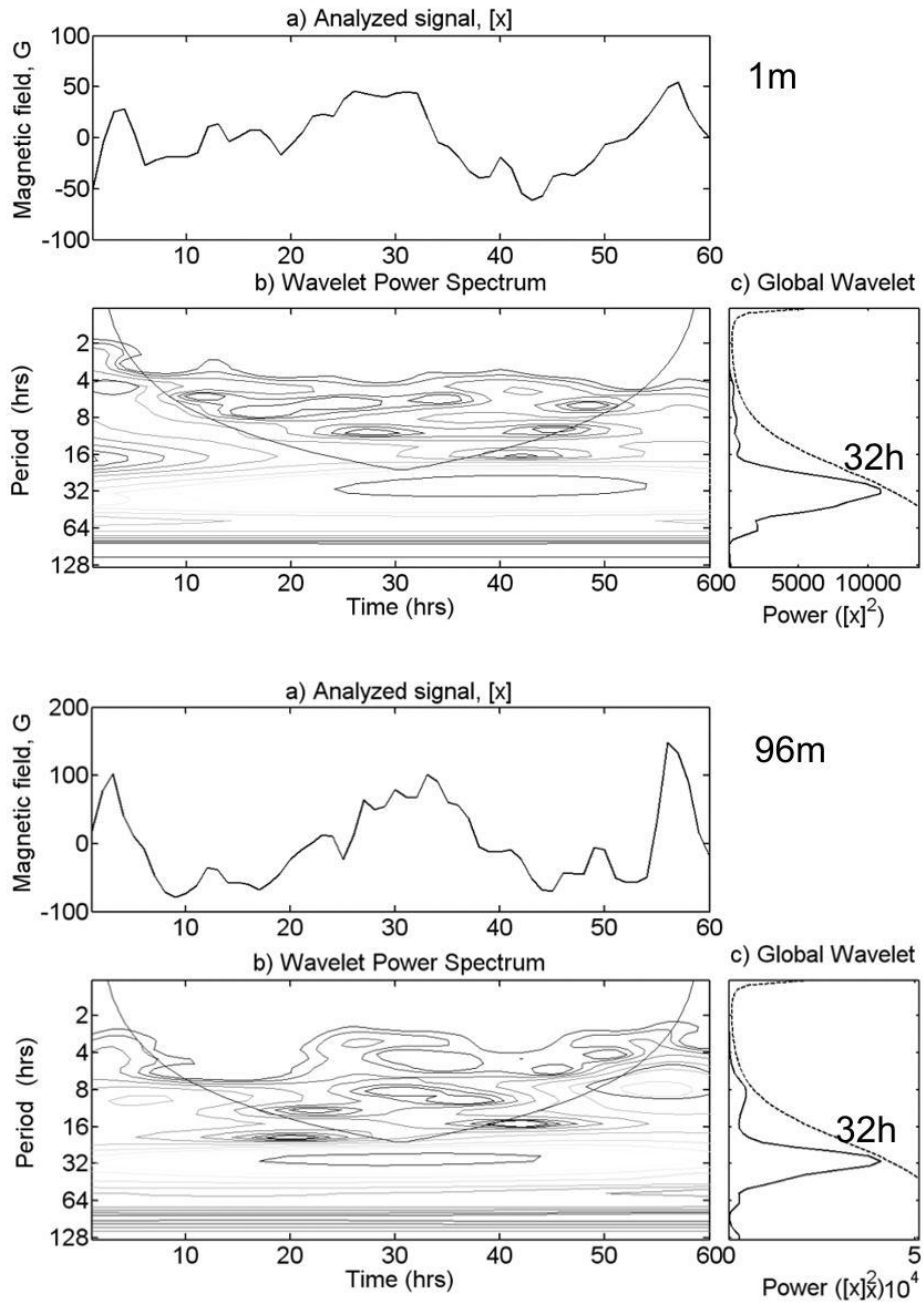

Figure 9. Another example of independence of the low-frequency part of sunspot power spectrum on the cadence of magnetograms (8 - 10 October 1999).

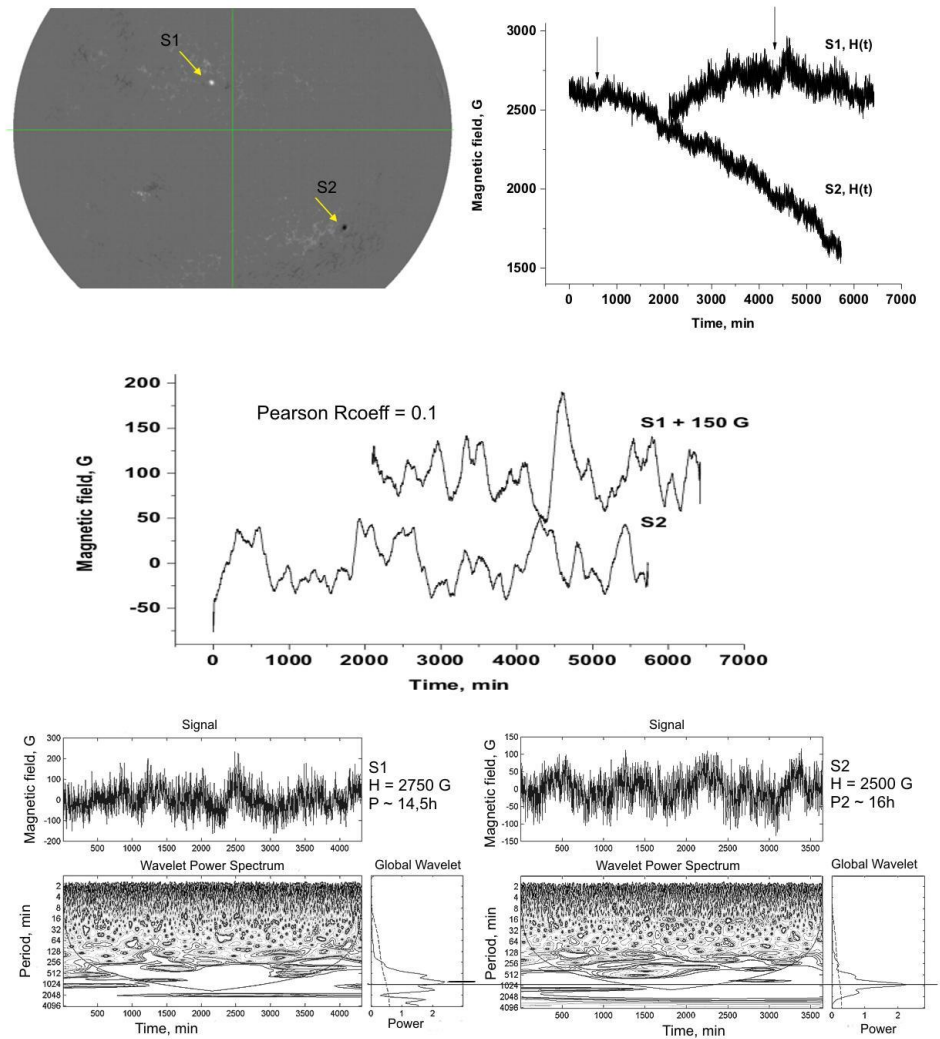

Figure 10. The results of processing of simultaneous magnetograms for two sunspots from different hemispheres, 24 - 26 September 1999.

As an example of this experiment, we present in Figures 7a and 7b the results of the treatment of magnetograms for two spatially separated sunspots [S1 and S2] for 24 - 26 September 1999 and 1 - 5 April 2002. Time series of one-minute cadence were used in the first case and the magnetograms with 96-minute cadence in the second. We selected the sunspots to be of opposite ( first case) and of the same polarity (second case), with several different magnetic strengths, that are located in different active regions (and even in different hemispheres of the Sun, as in the first example in Figure 7a). There is no physical connection between these sunspots. The oscillations on the eigenmodes are clearly observed in each of the spots. In the first case (Figure 7a), periods of 14.5 and 16 hours are visible, in accordance with a slight difference in their magnetic fields.

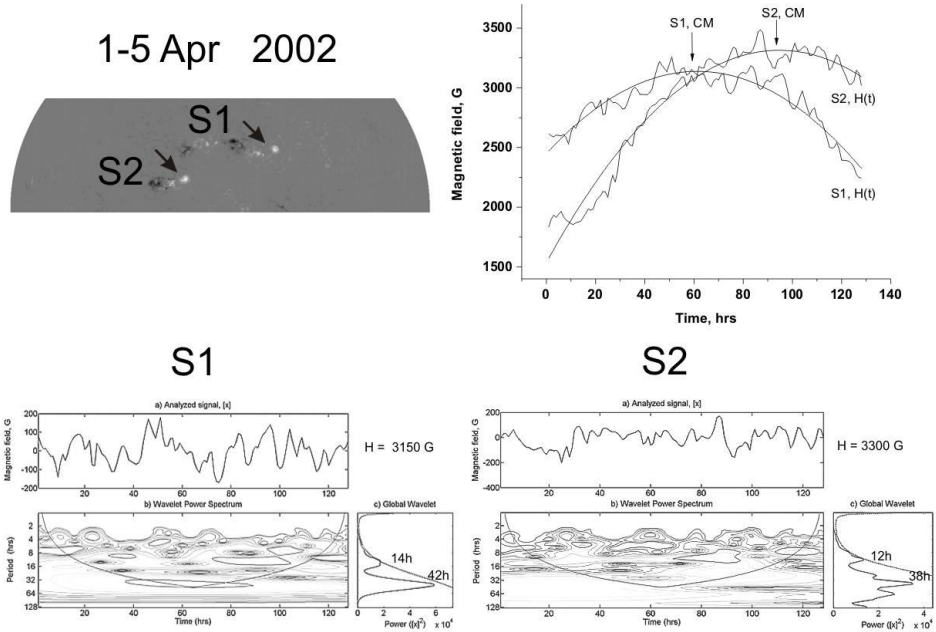

Figure 11. The results of processing of synchronous magnetograms for two sunspots from different groups on 1 - 5 April 2002.

In the second case (Figure 7b), another low-frequency mode with a period of about 38 - 42 hours is added to this mode. In both cases, there is no correlation of the oscillations in the different sunspots. The Pearson correlation coeficient is lower than 0.1. As is expected, the widely separated sunspots oscillate independently. No long-period artifacts were revealed in this study.

6. Conclusions

The examples given and discussed above clearly show the reliability of the results obtained with our technique of stabilizing a sunspot image; they clearly reveal the existence of long-period oscillations of the magnetic field in sunspots caused by the oscillation of the sunspot as a whole.

This clear picture of the oscillations was possible to obtain because the signal-to-noise ratio is sufficiently high in the original magnetograms.

i) The low-frequency eigenmodes of sunspots and surrounding magnetic elements are revealed both by ground-based and space observations. Studying these oscillations opens up many new possibilities to probe of physical parameters of active solar structures.

ii) The limiting low-frequency mode of the sunspot magnetic- field oscillations is the mode with the period of 12 - 32 hours). This period depends mainly on the magnetic- field strength of a sunspot.

iii) Except for the limiting eigenmode, in oscillatory spectra of sunspots the other (higher) modes are steadily present with periods of 40 - 45, 60 - 80, 135 - 170, 220 - 240, and 480 - 520 minutes. The amplitudes of these modes decrease monotonically and rapidly with increasing frequency. This is the characteristic feature of the overtones that arise because of the nonlinear character of the oscillations.

iv) The limiting eigenmode M1 is steadily maintained in sunspots throughout the observation period (five - seven days), but its amplitude normally has cyclic temporal variations (increase) with a period of one - two days, which coincides with the average lifetime of the supergranulation cells. Owing to these quasi-periodic perturbations of the M1 amplitude, the lower harmonic [M2] with its period of 2100 - 2880 minutes (35 - 48 hours) can be often found in the power spectrum of sunspots. It is not an eigenmode of the sunspot oscillations, because its period does not depend on the magnetic-field strength of the sunspot. Perhaps it arises as result of quasi-periodic perturbations of a sunspot, caused by external force (the emergence and decay of supergranulation cells surrounding the sunspot).

It is important to note that the main results of this work have been obtained by treating MDI magnetograms with one-minute cadence, which are definitely free from the influence of 12- and 24-hour artifacts that might have been caused by the peculiarities of telemetry transmission. More magnetograms with a 96-minute cadence were used to increase the statistical significance of results.

The authors consider the obtained results as preliminary, requiring additional studies, which may be stimulated by this work.

Acknowledgements We thank P.H. Scherrer and the SOHO team for providing the SOHO/MDI data. The work was supported by grant RFBR 13-02-00714, and by the Programs of the Russian Academia of Science: P-21 and P-22. Wavelet software was provided by C. Torrence and G. Compo, 1998 (available at paos.colorado.edu/research/wavelets/).

References

Bakunina, I.A., Abramov-Maximov, V.E., Lesovoy, S.V., Shibasaki, K., Solov'ev, A.A., Tikhomirov, Yu.V.: 2009a, In: Gopalswamy, N., Webb, D.F. (eds.), Universal Heliophysical Processes, Proc. Internat. Astron. Union, Symp. 257, Cambridge Univ. Press, 155. DOI. Bakunina, I.A., Solovyev, A.A., Nagovitsyn, Yu.A., Tikhomirov, Yu.V., Bakunin, V.L., Prosovetsky, D.V., Kuznetsova, S.M.: 2009b, Geomag. Aeron. 49, 1087. DOI.

Beckers, J.M., Schultz, R.B.: 1972, Solar Phys. 27, 61. ADS, DOI.
Chorley, N., Hnat, B., Nakariakov, V.M., Inglis, A.R., Bakunina, I.A.: 2010, Astron. Astrophys., 513, id. A27. DOI.
Domingo, V., Fleck, B., Poland, A.I.: 1995, Solar Phys. 162, 1. DOI
Efremov, V.I., Parfinenko, L.D., Solov'ev, A.A.: 2007, Astron. Rep. 51, 401. DOI
Efremov, V.I., Parfinenko, L.D., Solov'ev, A.A.: 2009, Cosmic Res. 47, 279. DOI.
Efremov, V.I., Parfinenko, L.D., Soloviev, A.A.: 2010, Solar Phys. 267, 279. DOI.
Efremov, V.I., Parfinenko, L.D., Solov'ev, A.A.: 2012, Cosmic Res. 50, 44. DOI.
Gelfreikh, G.B., Nagovitsyn Yu.A., Nagovitsyna E.Yu.: 2006, Pub. Astron. Soc. Japan 58, 29.
Kallunki, J., Riehokainen, A.: 2012, Solar Phys. 280, 347. DOI.
Koen, C.,: 1990, Astrophys. J. 348, 700. DOI.
Leibacher, J.W., Stein, R.F.: 1971, Astrophys. Let. 7, 191.
Leighton, R.B., Noyes, R.W., Simon, G.W.: 1962, Astrophys. J. 135, 474. DOI.
Pesnell, W.D., Thompson, B.J, Chamberlin, P.C.: 2012, Solar Phys. 275, 3. DOI
Scherrer, P.H., Bogart, R.S., Bush, R.I., Hoeksema, J.T., Kosovichev, A.G., Schou, J., Rosenberg, W., Springer, L., Tarbell, T.D., Title, A., Wolfson, C.J., Zayer, I., MDI Engineering Team: 1995, Solar Phys. 162, 129. DOI.
Scherrer, P.H., Schou, J., Bush, R.I., Kosovichev, A.G., Bogart, R.S., Hoeksema, J.T., Liu, Y., Duvall, T.L. Jr., Zhao, J., Title, A.M., Schrijver, C.J., Tarbell, T.D., Tomczyk, S.: 2012, Solar Phys. 275, 207. DOI.
Smirnova, V., Riehokainen A., Ryzhov V., Zhiltsov A., Kallunki J.: 2011, Astron. Astrophys. 534, id.A137. DOI
Smirnova, V., Riehokainen A., Solov'ev, A.A., Zhiltsov A., Kallunki J.: 2013a, Astron. Astrophys. 552, id.A23. DOI
Smirnova, V., Efremov, V.I., Parfinenko, L.D., Riehokainen A., Solov'ev, A.A.: 2013b, Astron. Astrophys. 554, id.A121. DOI
Solov'ev, A.A., Kirichek, E.A.: 2008, Astrophys. Bull.. 63. 169. DOI
Solov'ev, A.A., Kirichek, E.A.: 2009, Astron. Rep. 53 , 675. DOI Stahn,
T., Gizon, L.: 2008, Solar Phys. 251. 31. DOI
Torrence, C., Compo, G.P.: 1998, Bull. Am. Meteof. Soc. 79, 61.